\documentclass[a4paper,prb,twocolumn,showpacs]{revtex4}
\renewcommand{\vec}[1]{\mathbf{#1}}
\usepackage{graphicx}
\begin{document}

\title{Response functions and superfluid density in a weakly
  interacting Bose gas with non-quadratic dispersion.}
\author{Jonathan Keeling} \affiliation{Department of Physics, Massachusetts
  Institute of Technology, 77 Mass. Ave., Cambridge, MA 02139, USA}

\begin{abstract}
  Motivated by the experimental search for Bose condensation of
  quasiparticles in semiconductors, the response functions of a weakly
  interacting Bose gas, with isotropic but non-quadratic dispersion, are
  considered.  Non-quadratic dispersion modifies the definition of
  particle current, and leads to modified sum rules for the
  current-current response function.  The effect of these
  modifications on the Berezhinski-Kosterlitz-Thouless transition is
  discussed.
\end{abstract}
\maketitle

Recently, there has been increasing interest in Bose condensation of
quasiparticles in solid state systems.
Examples include indirect excitons in semiconductor quantum
wells\cite{butov04}, exciton-polaritons in semiconductor
microcavities\cite{dang98,deng02,richard05}, quantum hall bilayer
excitons\cite{eisenstein04:nature,eisenstein04:science}, and spin
``triplons'' in copper compounds\cite{ruegg03,jaime04,radu05}.
In many of these cases, the composite nature of the quasiparticle
leads to significant deviations from a quadratic dispersion.
Such deviations mean that a current defined by $\vec{J}(x)
=\psi^{\dagger}(x) i \nabla \psi^{}(x)$ is no longer correct: such a
current is not conserved, and so its correlation functions do not obey
simple sum rules.
Neither can this problem be extricated by working in terms of more
fundamental fields, e.g.\ the photon/exciton fields for the polariton
problem, as in such an example the photon current is not conserved, the
Hamiltonian has terms by which photon current is transfered to exciton
current and back again.

There is an obvious solution to this problem: the correct definition
of current is the Noether current associated with invariance of the
action under global phase rotations.
Such a definition automatically leads to a conserved current, which
for quadratic dispersion is just the standard definition.
The definition of current and its response functions are of particular
importance due to another common feature of these solid state systems
in which condensation is sought: they are two dimensional, and
therefore the transition is of the Berezhinski-Kosterlitz-Thouless
(BKT) class \cite{kosterlitz73,nelson77}.
Therefore, to find the critical conditions at which the transition
should occur, it is necessary to find the superfluid stiffness,
including effects of depletion by density fluctuations.
This is most naturally achieved by finding the current response
functions, and thus separating the current response into normal and
superfluid components\cite{nelson83,fisher88,stoof93}.
For the weakly interacting case, one may then perturbatively
evaluate the current response functions.
Such a perturbative evaluation relies on two properties of the current
response: a sum rule on the longitudinal response function (a
consequence of using a conserved current)\cite{pitaevskii03}, and an
understanding of the effect of vertex corrections on the transverse
response functions\cite{takahashi57,ward50,griffin94}.

The aim of this article is to discuss the correct generalisation of
response functions for non-quadratic, but isotropic, quasi-particle
dispersion.
Previous work on the BKT transition in a model of weakly interacting
bosons with non-quadratic dispersion\cite{kavokin03,malpuech03} 
did not generalise the current in this manner.
As a result, the current in that work was not conserved, and so there
is no sum rule relating the longitudinal response function to density.
As the spectrum considered there was quadratic for small momenta, any
formalism which recovers the standard form at low densities (i.e.\ when
only low momentum particles are excited) will agree.
However, at higher densities, when particles beyond the quadratic
dispersion contribute to the current response, there are differences
between the method described here, and the method in those previous
works, as will be shown below.

To be precise, consider the following model of a weakly interacting
Bose gas:
\begin{equation}
  \label{eq:1}
  H =  
  \sum_\vec{k} \epsilon_\vec{k} \psi^{\dagger}_\vec{k} \psi^{}_\vec{k}
  +
  \frac{g}{2}
  \sum_{\vec{k},\vec{k}^{\prime},\vec{q}} 
  \psi^{\dagger}_{\vec{k}+\vec{q}} \psi^{\dagger}_{\vec{k}^{\prime} - \vec{q}}
  \psi^{}_{\vec{k}^{\prime}} \psi^{}_\vec{k},
\end{equation}
where $\epsilon_\vec{k}$ is isotropic, and has a quadratic part as $k
\to 0$, but is otherwise general.
This Hamiltonian is invariant under global phase rotations, and so
there is an associated Noether current $\vec{J}$ given
by\cite{nagaosa}:
\begin{equation}
  \label{eq:2}
  J_i(x) = 
  \frac{\delta S}{\delta \partial_i \psi^{\dagger}(x)}
  i \psi^{\dagger}(x)
  -
  \frac{\delta S}{\delta \partial_i \psi^{}(x)}
  i \psi^{}(x),
\end{equation}
where $S$ is the action from the Hamiltonian in eq.~(\ref{eq:1}).
By definition, this current is conserved, so:
\begin{equation}
  \label{eq:3}
  \left[ H, \rho(\vec{q}) \right] = \vec{q}\cdot\vec{J}(\vec{q}),
  \qquad
  \rho(\vec{q}) = 
  \sum_\vec{k} \psi^{\dagger}_{\vec{k} + \vec{q}/2} \psi^{}_{\vec{k} - \vec{q}/2}.
\end{equation}
In the following, we will be interested in the static response of the
system to an applied force that couples to such a current, described
by the response function:
\begin{equation}
  \label{eq:4}
  \chi_{ij}(\omega=0,\vec{q})
  =
  2 \int_0^{\beta} d \tau
  \left<\left<
    J_i(\vec{q},\tau) J_j(-\vec{q},0)
  \right>\right>,
\end{equation}
where double angle brackets indicate quantum and thermal averaging.
For an isotropic system, the most general form of the response function
is:
\begin{equation}
  \label{eq:5}
  \chi_{ij}(\vec{q}) = 
  \chi_T(q) \left( \delta_{ij} - \frac{q_i q_j}{q^2} \right)
  +
  \chi_L(q) \frac{q_i q_j}{q^2}.
\end{equation}
The standard rotating bucket argument still applies in dividing
the response into a superfluid part that contributes only to
$\chi_L$ and a normal part that contributes to both $\chi_L$ and
$\chi_T$.
With a quadratic dispersion, the relevant quantity is $\rho_s/m =
\lim_{q \to 0}\left[  \chi_L(q) - \chi_T(q) \right]$.
With non-quadratic dispersion mass is now q dependent, so the
identification of $\rho_s$ and $m$ separately is not possible, but it
is not necessary; the effective vortex action depends only on the well
defined quantity:
\begin{equation}
  \label{eq:6}
  \chi_s = \lim_{q \to 0} [ \chi_L(q) - \chi_T(q) ].
\end{equation}
Since the current used is by definition conserved, $\chi_L$ will be
subject to a sum rule; a generalisation of the sum rule that would for
quadratic dispersion relate $\chi_L$ to the density.
Below, this sum rule is evaluated, and thus $\chi_L$ and $\chi_T$ are
calculated in a perturbative expansion.


Before evaluating this sum rule, it is first worth stressing why the
above generalisation gives the quantity appropriate to the BKT
transition.
The BKT transition is associated with the unbinding of vortex pairs.
The conditions at which the transition occurs are therefore described
by the effective vortex-vortex interaction, and the vortex fugacity.
Starting from a microscopic model, these quantities both depend on the
phase stiffness: the coefficient of $(\nabla \phi(x))^2 \sim q^2
\phi_q^2$ in the effective action.
It is only this quadratic phase dispersion which matters:
non-quadratic terms in the phase dispersion lead only to short range
vortex interactions, while the quadratic term leads to a logarithmic
confining potential.
However, the phase stiffness is modified by density fluctuations, and
the non-quadratic dispersion of density fluctuations can modify the phase
stiffness.
Non-quadratic dispersion matters because after integrating out density
fluctuations, non-quadratic dispersion of density fluctuations modifies
the coefficient of quadratic dispersion of phase fluctuations.
It is technically easier to evaluate the current response functions
than to directly integrate out density fluctuations, and the
associated definitions of superfluid stiffness are
equivalent\cite{griffin94}.

The sum rule for $\chi_L(q) = q_i q_j \chi_{ij}(\vec{q}) /q^2$ follows from
eq.~(\ref{eq:4}) and eq.~(\ref{eq:3}), and the standard procedure, as
described for example in ref.\onlinecite{pitaevskii03}:
\begin{eqnarray}
  \label{eq:7}
  \chi_L(q) 
  &=&
  \frac{2}{q^2}\int_0^{\beta} d \tau
  \left<\left<
    \vec{q}\cdot\vec{J}(\vec{q},\tau) \vec{q}\cdot\vec{J}(-\vec{q},0)
  \right>\right>
  \nonumber\\
  &=&
  \frac{1}{\mathcal{Z}q^2}
  \sum_n e^{-\beta E_n}
  \left< n \right|
  \left[ \rho(\vec{q}), \vec{q}\cdot\vec{J}(\vec{q}) \right]
  \left| n \right>,
\end{eqnarray}
where one use has been made of the commutation relation
eq.~(\ref{eq:3}).
Writing the commutation relations explicitly in terms of the
$\psi^{\dagger},\psi$ operators, one has:
\begin{eqnarray*}
  \vec{q}\cdot\vec{J}(\vec{q})
  \!\!&=&\!\!
  \sum_{\vec{k}}
  \left(
    \epsilon_{\vec{k} + \vec{q}/2} - \epsilon_{\vec{k} - \vec{q}/2}
  \right)
  \psi^{\dagger}_{\vec{k} + \vec{q}/2}
  \psi^{}_{\vec{k} - \vec{q}/2},
  \\
  \left[ \rho(\vec{q}), \vec{q}\cdot\vec{J}(\vec{q}) \right]
  \!\!&=&\!\!
  \sum_{\vec{k}}
  \left(
    \epsilon_{\vec{k} + \vec{q}}+ \epsilon_{\vec{k} - \vec{q}} - 2\epsilon_{\vec{k}}
  \right)
  \psi^{\dagger}_{\vec{k} + \vec{q}/2}
  \psi^{}_{\vec{k} - \vec{q}/2}.
\end{eqnarray*}
In the limit $q\to 0$, the terms in parentheses are independent of the
direction of $k$, so we may average over solid angles $d \Omega$ and
thus have:
\begin{eqnarray}
  \label{eq:8}
  \lim_{q\to 0} \chi_L(q) 
  &=&
  \sum_\vec{k} g_k
  \left<\left<
      \psi^{\dagger}_{\vec{k}}
      \psi^{}_{\vec{k}}
    \right>\right>,
  \\
  g_k
  \label{eq:9}
  &=&
  \int \frac{d\Omega}{\Omega}
  \lim_{q\to 0}
  \left(
    \frac{
      \epsilon_{\vec{k} + \vec{q}}+ \epsilon_{\vec{k} - \vec{q}} - 2\epsilon_{\vec{k}}
    }{q^2}
  \right).
\end{eqnarray}
Since dispersion is isotropic, we may write it as $\epsilon_\vec{k} =
f(k^2)$, thus $g_k = (4/d)k^2 f^{\prime\prime}(k^2) + 2
f^{\prime}(k^2)$, and as expected a quadratic dispersion, $f(x)=x/2m$
reduces to $g_k = 1/m$.
In general, eq.~(\ref{eq:8}) can be considered as density weighted by
effective inverse mass of each momentum component.
The longitudinal response is thus reduced to finding an approximation
scheme for the occupation of each finite $k$ mode, which will be
discussed below.

As yet we have only written explicitly the longitudinal component
of the current.  
To find correlations of the transverse component, it is convenient to
write:
\begin{equation}
  \label{eq:10}
  J_i(\vec{q}) = 
  \sum_\vec{k} 
  \Psi^{\dagger}_{\vec{k} + \vec{q}}
  \gamma_i(\vec{k} + \vec{q}, \vec{k})
  \Psi^{}_{\vec{k}}
  , \quad
  \Psi^{}_{\vec{k}}
  =
  \left(
    \begin{array}{c}
      \psi_{\vec{k}} \\ \psi^{\dagger}_{-\vec{k}}
    \end{array}
  \right),
\end{equation}
and from conservation of current, we have:
\begin{equation}
  \label{eq:11}
  q_i \gamma_i(\vec{k} + \vec{q}, \vec{k})
  =
  \sigma_3 \left(\epsilon_{\vec{k}+\vec{q}} - \epsilon_{\vec{k}}\right).
\end{equation}
Thus, we know the projection of the vector $\gamma_i$ onto one axis;
what remains is to find its direction.
This follows from the definition in eq.~(\ref{eq:2}), which shows that
under the interchange $\psi^{}_{\vec{k}} \leftrightarrow \psi^{\dagger}_{\vec{k}}$
the current changes as $J_i \to - J_i$.
With a little algebra, it can be seen that this directly implies
$\gamma_i(\vec{p},\vec{q}) = (p_i + q_i) f(\vec{p},\vec{q}) \sigma_3$
where $f(\vec{p},\vec{q})$ is a scalar function chosen to satisfy
eq.~(\ref{eq:11}).

\begin{figure}[htpb]
  \centering
  \raisebox{26pt}{$\Gamma_i(\vec{k}+\vec{q}, \vec{k}) = \gamma_i(\vec{k}+\vec{q}, \vec{k}) + $}
  \begin{picture}(120,60)
    \put(0,0){\includegraphics{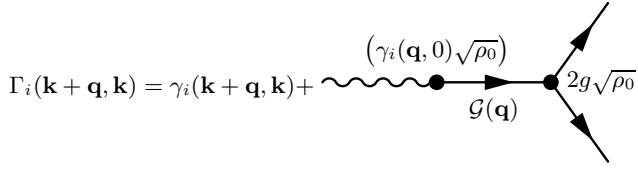}}
    \put(64.79951,24){\makebox(0,0)[t]{{$\mathcal {G}(\vec {q})$}}}
    \put(43.1997,36){\makebox(0,0)[b]{{$\left ( \gamma _i(\vec {q},0) \sqrt {\rho _0} \right )$}}}
    \put(92.39941,30){\makebox(0,0)[l]{{$2g\sqrt {\rho _0}$}}}
  \end{picture}
  \caption{Vertex corrections required at one-loop order.}
  \label{fig:vertex}
\end{figure}

From this definition of current, it is now possible to calculate 
the current response function perturbatively.
As in the quadratic dispersion case, the leading order perturbative
calculation relies on properties of the corrections to the current
vertex $\gamma_i$ that result from interactions.
For clarity, the standard argument\cite{pitaevskii03,griffin94} is
summarised here.
A full calculation of the response would be given by
\begin{equation}
  \label{eq:12}
  \lim_{q\to 0}
  \chi_{ij}(q) = 2 \int \frac{d^d k}{(2\pi)^d} \mathrm{Tr} \left(
    \mathcal{G}(\vec{k}) 
    \Gamma_i(\vec{k}, \vec{k})
    \mathcal{G}(\vec{k})
    \gamma_j(\vec{k}, \vec{k})
  \right),
\end{equation}
where $\mathcal{G}(\vec{k})$ is the Green's function in the Nambu
representation indicated in eq.~(\ref{eq:10}), and $\Gamma_i$ is
the vertex $\gamma_i$ including corrections.
At one loop order, vertex corrections are necessary to make
eq.~(\ref{eq:12}) satisfy the sum rule, eq.~(\ref{eq:8}).
However, it can be seen that these required vertex corrections are of
the form shown in Fig.~\ref{fig:vertex}.
Since these involve a vertex where current couples directly to
the condensate, they involve $\gamma_i(\vec{q},0)$, which, due
to the previous discussion of the direction of $\gamma_i$, is
proportional to $q_i$.
Such a correction therefore only changes the longitudinal response.
Therefore, we may safely evaluate the transverse response at
one-loop order without such corrections.

Having understood why vertex corrections can be ignored, the
perturbative calculation of $\chi_T$ now follows directly; writing:
\begin{equation}
  \label{eq:13}
  \gamma_i(\vec{k},\vec{k}) 
  = 
  k_i
  \lim_{q\to 0} 
  \left(
    \frac{%
      \epsilon_{\vec{k}+\vec{q}/2} - \epsilon_{\vec{k}-\vec{q}/2}
    }{\vec{k}\cdot\vec{q} } 
  \right)
  \sigma_3
  =
  2 k_i f^{\prime}(k^2),
\end{equation}
with $f(k^2)=\epsilon_{\vec{k}}$ as before for an isotropic mass,
we thus have:
\begin{equation}
  \label{eq:14}
  \lim_{q \to 0} \chi_T(q)
  =
  \frac{2}{d} 
  \int \frac{d^d k}{(2\pi)^d} 
  k^2 (2 f^{\prime}(k^2))^2
  \mathrm{Tr} \left(
    \mathcal{G}(\vec{k}) 
    \sigma_3
    \mathcal{G}(\vec{k})
    \sigma_3
  \right).
\end{equation}
To complete the evaluation of $\chi_T$ then requires
an explicit form for the Green's function.
The Bogoliubov one-loop approximation for the condensed Green's
function is:
\begin{equation}
  \label{eq:15}
  \mathcal{G}(\omega,\vec{k})
  = 
  \frac{1}{\omega^2 + \xi_k^2}
  \left(
    \begin{array}{cc}
      -i\omega + \epsilon_{k} + \mu & - \mu \\
      -\mu & i\omega + \epsilon_{k} + \mu
    \end{array}
  \right),
\end{equation}
where $\xi_k=\sqrt{\epsilon_{k}(\epsilon_{k} + 2\mu)}$ is
the Bogoliubov quasiparticle energy.
Thus, eq.~(\ref{eq:14}) becomes:
\begin{eqnarray}
  \label{eq:16}
  \mathrm{Tr}
  \left(
    \mathcal{G} \sigma_3 \mathcal{G} \sigma_3
  \right)
  &=&
  \frac{1}{2} \sum_{\omega_n} \frac{
    \xi_k^2 - \omega_n^2
  }{(\omega_n^2 + \xi_k^2)^2}
  = -\frac{\beta}{2} n_B^{\prime}(\xi_k),
  \\
  \label{eq:17}
  \lim_{q\to 0} \chi_T(q)
  &=&
  -
  \frac{2}{d} 
  \int \frac{d^d k}{(2\pi)^d} 
  \frac{k^2 (2 f^{\prime}(k^2))^2}{2}
  n_B^{\prime}(\xi_k).
\end{eqnarray}

Finally, to find $\chi_L$ requires evaluation of the average occupation
of each $k$ mode in eq.~(\ref{eq:8}).
In evaluating eq.~(\ref{eq:8}), as the effects of fluctuations in the
presence of a condensate are required, the condensate depletion due to
fluctuations must be included in order to derive a consistent answer
\cite{keeling05,khawaja02}.
This means $\rho_0=\left<\left< \psi^{\dagger}_0 \psi^{}_0 \right>
\right>$ must include fluctuation corrections, determined by
considering a chemical potential coupled only to $k=0$ modes, or
equivalently by using the Hugenholtz-Pines relation at one loop order:
\begin{equation}
  \label{eq:18}
  \rho_0 
  = 
  \frac{\mu}{g} - 
  \sum_{\vec{k}} \left[ 
    2 \left<\!\left< \psi^{\dagger}_\vec{k} \psi^{}_\vec{k} \right>\!\right>
    + \frac{1}{2}
    \left(
      \left<\!\left< \psi^{\dagger}_\vec{k} \psi^{\dagger}_\vec{k} \right>\!\right>
      +
      \mathrm{h.c.}
    \right) \right].
\end{equation}
Inserting this in eq.~(\ref{eq:8}) yields the final form:
\begin{eqnarray}
  \label{eq:19}
  \lefteqn{\lim_{q\to 0}\chi_L(q)
    =
  g_0 \frac{\mu}{T_{2B}}}
  \nonumber\\
  &&-
  \int \frac{d^dk}{(2\pi)^d} \left[
    n_B(\xi_k) \left( 
      g_0 \frac{2\epsilon_k + \mu}{\xi_k} - g_k \frac{\epsilon_k+\mu}{\xi_k}
    \right)
  \right.
  \nonumber\\
  &&+
  \left.
    (2g_0 - g_k) \frac{\epsilon_k + \mu - \xi_k}{2\xi_k}
    +
    \frac{g_0 \mu}{2}
    \frac{\xi_k - \epsilon_k - \mu}{\xi_k(\epsilon_k+\mu)}
  \right].
\end{eqnarray}
(To avoid the ultra-violet divergence associated with a delta-function
interaction, the standard T-matrix renormalisation\cite{khawaja02} has
been performed, thus $T_{2B}$ is the renormalised two-body T-matrix
corresponding to the bare interaction $g$).

\begin{figure}[htpb]
  \centering
  \includegraphics[width=\columnwidth]{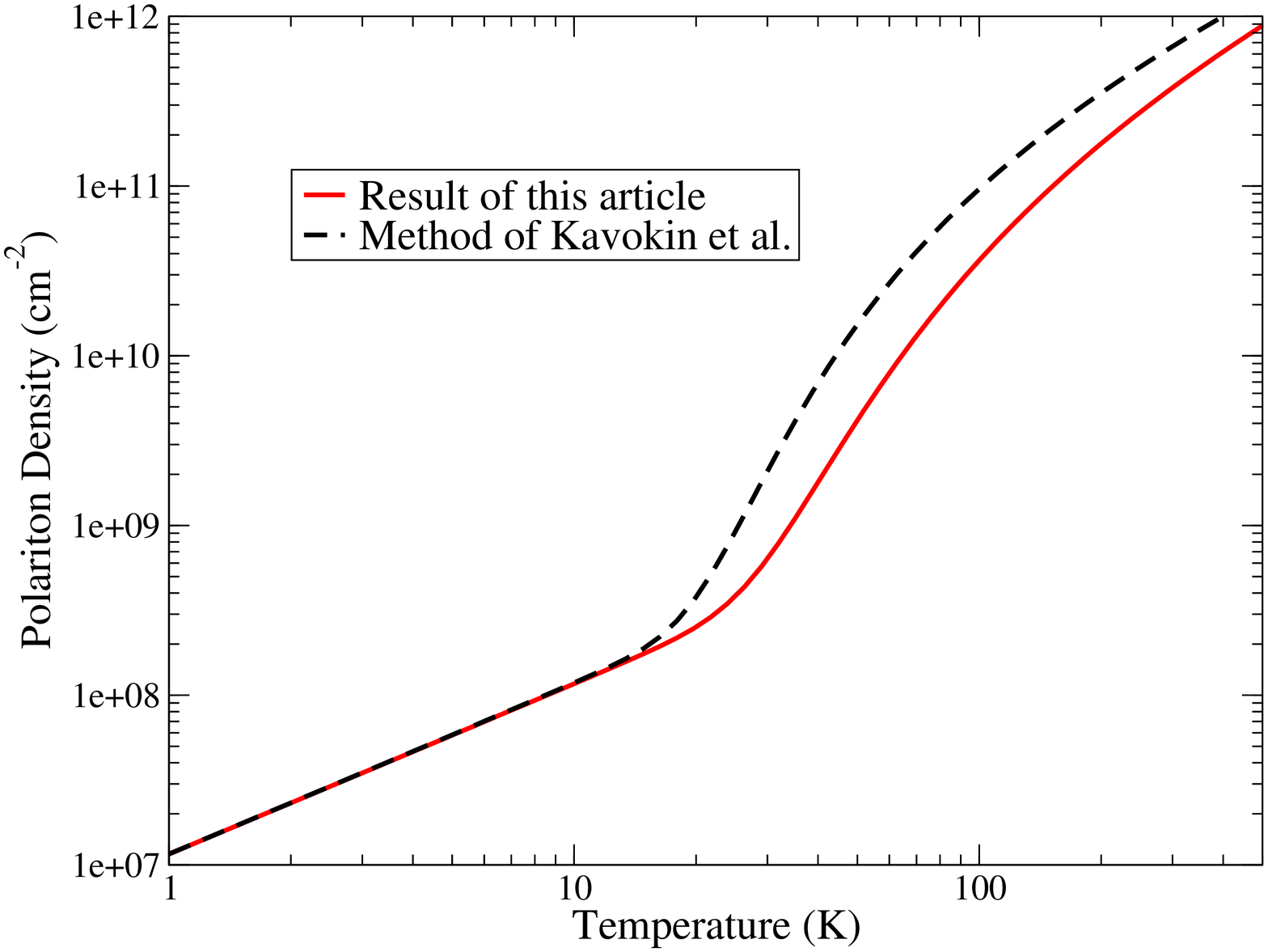}
  \caption{Comparison of calculation of BKT critical density vs
    temperature for the method discussed here and the method in
    ref.~\onlinecite{kavokin03,malpuech03}.  At low densities, both
    calculations agree, as non-quadratic effects are irrelevant, but
    where such effects matter, their predictions differ.}
  \label{fig:compare}
\end{figure}

In two dimensions, the BKT transition is found, as discussed above, by
evaluating eq.~(\ref{eq:17}) and eq.~(\ref{eq:19}) at a fixed
temperature, and finding the value of $\mu$ which satisfies
$\lim_{q \to 0}\left[ \chi_L(q) - \chi_T(q) \right] = \pi k_B T/2$.
The discussion up to now has been for a generic isotropic dispersion.
For illustration, Fig.~\ref{fig:compare} shows the results of such a
calculation with a dispersion appropriate to exciton-polaritons:
\begin{equation}
  \label{eq:20}
  f(k^2) = \frac{1}{2} \left[ 
    \frac{k^2}{2m_1}
    - 
    \sqrt{\left( 
        \frac{k^2}{2m_2}
      \right)^2
      + 
      \Omega_R^2}
    \right],
\end{equation}
with $1/m_1 = 1/m_X + 1/m_P$, and $1/m_2 = 1/m_X - 1/m_P$. 
Parameters are chosen close to those of the experiments of
ref.~\onlinecite{richard05} in CdTe: exciton mass $m_X=0.08 m_\mathrm{e}$ ,
photon mass $m_P=2.58\times10^{-5} m_\mathrm{e}$, $\Omega_R=26
\mathrm{meV}$, and $T_{2B} = 13 \mathrm{meV}/10^{11} \mathrm{cm}^{-2}$.

In the normal state, the transverse and longitudinal response
functions should become equal.
It is instructive to see how the expression for effective mass,
weighting the density, appears in such a calculation.
In the normal state, there are no condensate depletion effects
to worry about, and so:
\begin{equation}
  \label{eq:21}
  \chi_L(0)
  =
  \int \frac{d^d k}{(2\pi)^d}
  g_k 
  n_B(f(k^2)),
\end{equation}
The one-loop transverse response is as in eq.~(\ref{eq:17}), but with
$\xi_k \to \epsilon_k = f(k^2)$.
To see that they agree, it is convenient to rewrite eq.~(\ref{eq:17})
with a change of integration variables.
We first introduce $x=k^2$, so $d^dk = S_d x^{d/2-1} dx/2$, with $S_d$
the surface of the d-dimensional hypersphere, and then change
integration variable again to $f(x)$, with $dx = df /f^{\prime}(x)$,
giving:
\begin{eqnarray}
  \label{eq:22}
  \chi_T(0)
  &=& 
  - \frac{2}{d} 
  \int \frac{S_d df}{(2\pi)^d} 
  \frac{x^{d/2 -1} }{2  f^{\prime}(x)} 
  \frac{x (2 f^{\prime}(x))^2}{2} \frac{d n_B(f)}{df}
  \nonumber\\
  &=&
  \frac{2}{d}
  \int \frac{S_d df}{(2\pi)^d} 
   \left( 
     \frac{1}{2 f^{\prime}(x)} \frac{d}{dx}
     \left[ x^{d/2} 2 f^{\prime}(x) \right]
   \right)
   n_B(f)
   \nonumber\\
   &=&
   \frac{2}{d} \frac{d}{2} 
   \int \frac{d^d k}{(2\pi)^d}
   g_k
   n_B(f(k^2)),
\end{eqnarray}
where the second line is integration by parts, and the last used
$\partial_x(2 x^{d/2} f^{\prime}(x)) = (d/2) x^{d/2 -1} g(x)$.

In conclusion, a formalism for calculating the transverse and
longitudinal response functions of a Bose gas with arbitrary isotropic
dispersion has been presented.
A sum rule relates the longitudinal response to density weighted by
effective inverse mass at a given momentum.
Using such a formalism recovers the equivalence of transverse
and longitudinal responses in the normal state.
This formalism allows a consistent formulation of the critical
conditions for the BKT transition in a two-dimensional Bose gas.

\begin{acknowledgments}
  I would like to thank F.~M.~Marchetti and M.~H.~Szymanska for the
  suggestion that lead to this work, and P.~B.~Littlewood for helpful
  discussion, and to acknowledge financial support from the Lindemann
  Trust.
\end{acknowledgments}


\begin{thebibliography}{23}
\expandafter\ifx\csname natexlab\endcsname\relax\def\natexlab#1{#1}\fi
\expandafter\ifx\csname bibnamefont\endcsname\relax
  \def\bibnamefont#1{#1}\fi
\expandafter\ifx\csname bibfnamefont\endcsname\relax
  \def\bibfnamefont#1{#1}\fi
\expandafter\ifx\csname citenamefont\endcsname\relax
  \def\citenamefont#1{#1}\fi
\expandafter\ifx\csname url\endcsname\relax
  \def\url#1{\texttt{#1}}\fi
\expandafter\ifx\csname urlprefix\endcsname\relax\def\urlprefix{URL }\fi
\providecommand{\bibinfo}[2]{#2}
\providecommand{\eprint}[2][]{\url{#2}}

\bibitem[{\citenamefont{Butov}(2004)}]{butov04}
\bibinfo{author}{\bibfnamefont{L.~V.} \bibnamefont{Butov}},
  \bibinfo{journal}{J. Phys.: Cond. Mat} \textbf{\bibinfo{volume}{16}},
  \bibinfo{pages}{R1577} (\bibinfo{year}{2004}).

\bibitem[{\citenamefont{Dang et~al.}(1998)\citenamefont{Dang, Heger, Andr\'{e},
  B{\oe}uf, and Romestain}}]{dang98}
\bibinfo{author}{\bibfnamefont{L.~S.} \bibnamefont{Dang}},
  \bibinfo{author}{\bibfnamefont{D.}~\bibnamefont{Heger}},
  \bibinfo{author}{\bibfnamefont{R.}~\bibnamefont{Andr\'{e}}},
  \bibinfo{author}{\bibfnamefont{F.}~\bibnamefont{B{\oe}uf}}, \bibnamefont{and}
  \bibinfo{author}{\bibfnamefont{R.}~\bibnamefont{Romestain}},
  \bibinfo{journal}{Phys. Rev. Lett.} \textbf{\bibinfo{volume}{81}},
  \bibinfo{pages}{3920} (\bibinfo{year}{1998}).

\bibitem[{\citenamefont{Deng et~al.}(2002)\citenamefont{Deng, Weihs, Santori,
  Bloch, and Yamamoto}}]{deng02}
\bibinfo{author}{\bibfnamefont{H.}~\bibnamefont{Deng}},
  \bibinfo{author}{\bibfnamefont{G.}~\bibnamefont{Weihs}},
  \bibinfo{author}{\bibfnamefont{C.}~\bibnamefont{Santori}},
  \bibinfo{author}{\bibfnamefont{J.}~\bibnamefont{Bloch}}, \bibnamefont{and}
  \bibinfo{author}{\bibfnamefont{Y.}~\bibnamefont{Yamamoto}},
  \bibinfo{journal}{Science} \textbf{\bibinfo{volume}{298}},
  \bibinfo{pages}{199} (\bibinfo{year}{2002}).

\bibitem[{\citenamefont{Richard et~al.}(2005)\citenamefont{Richard, Kasprzak,
  Romestain, Andr\'{e}, and Dang}}]{richard05}
\bibinfo{author}{\bibfnamefont{M.}~\bibnamefont{Richard}},
  \bibinfo{author}{\bibfnamefont{J.}~\bibnamefont{Kasprzak}},
  \bibinfo{author}{\bibfnamefont{R.}~\bibnamefont{Romestain}},
  \bibinfo{author}{\bibfnamefont{R.}~\bibnamefont{Andr\'{e}}},
  \bibnamefont{and} \bibinfo{author}{\bibfnamefont{L.~S.} \bibnamefont{Dang}},
  \bibinfo{journal}{Phys. Rev. Lett.} \textbf{\bibinfo{volume}{94}},
  \bibinfo{pages}{187401} (\bibinfo{year}{2005}).

\bibitem[{\citenamefont{Eisenstein and MacDonald}(2004)}]{eisenstein04:nature}
\bibinfo{author}{\bibfnamefont{J.~P.} \bibnamefont{Eisenstein}}
  \bibnamefont{and} \bibinfo{author}{\bibfnamefont{A.~H.}
  \bibnamefont{MacDonald}}, \bibinfo{journal}{Nature}
  \textbf{\bibinfo{volume}{432}}, \bibinfo{pages}{691} (\bibinfo{year}{2004}).

\bibitem[{\citenamefont{Eisenstein}(2004)}]{eisenstein04:science}
\bibinfo{author}{\bibfnamefont{J.~P.} \bibnamefont{Eisenstein}},
  \bibinfo{journal}{Science} \textbf{\bibinfo{volume}{305}},
  \bibinfo{pages}{950} (\bibinfo{year}{2004}).

\bibitem[{\citenamefont{R\"{u}egg et~al.}(2003)\citenamefont{R\"{u}egg,
  Cavadinin, Furrer, G\"{u}del, Kr\"{a}mer, Mukta, Wildes, Habicht, and
  P.Worderwisch}}]{ruegg03}
\bibinfo{author}{\bibfnamefont{C.}~\bibnamefont{R\"{u}egg}},
  \bibinfo{author}{\bibfnamefont{N.}~\bibnamefont{Cavadinin}},
  \bibinfo{author}{\bibfnamefont{A.}~\bibnamefont{Furrer}},
  \bibinfo{author}{\bibfnamefont{H.-U.} \bibnamefont{G\"{u}del}},
  \bibinfo{author}{\bibfnamefont{K.}~\bibnamefont{Kr\"{a}mer}},
  \bibinfo{author}{\bibfnamefont{H.}~\bibnamefont{Mukta}},
  \bibinfo{author}{\bibfnamefont{A.}~\bibnamefont{Wildes}},
  \bibinfo{author}{\bibfnamefont{K.}~\bibnamefont{Habicht}}, \bibnamefont{and}
  \bibinfo{author}{\bibnamefont{P.Worderwisch}}, \bibinfo{journal}{Nature}
  \textbf{\bibinfo{volume}{423}}, \bibinfo{pages}{62} (\bibinfo{year}{2003}).

\bibitem[{\citenamefont{Jaime et~al.}(2004)\citenamefont{Jaime, Correa,
  Harrison, Batista, Kawashima, Kazuma, Jorge, Stern, Heinmaa, Zvyagin
  et~al.}}]{jaime04}
\bibinfo{author}{\bibfnamefont{M.}~\bibnamefont{Jaime}},
  \bibinfo{author}{\bibfnamefont{V.~F.} \bibnamefont{Correa}},
  \bibinfo{author}{\bibfnamefont{N.}~\bibnamefont{Harrison}},
  \bibinfo{author}{\bibfnamefont{C.~D.} \bibnamefont{Batista}},
  \bibinfo{author}{\bibfnamefont{N.}~\bibnamefont{Kawashima}},
  \bibinfo{author}{\bibfnamefont{Y.}~\bibnamefont{Kazuma}},
  \bibinfo{author}{\bibfnamefont{G.~A.} \bibnamefont{Jorge}},
  \bibinfo{author}{\bibfnamefont{R.}~\bibnamefont{Stern}},
  \bibinfo{author}{\bibfnamefont{I.}~\bibnamefont{Heinmaa}},
  \bibinfo{author}{\bibfnamefont{S.~A.} \bibnamefont{Zvyagin}},
  \bibnamefont{et~al.}, \bibinfo{journal}{Phys. Rev. Lett.}
  \textbf{\bibinfo{volume}{93}}, \bibinfo{pages}{087203}
  (\bibinfo{year}{2004}).

\bibitem[{\citenamefont{Radu et~al.}(2005)\citenamefont{Radu, Wilhelm,
  V.Yushankhai, Kovrizhin, Coldea, Tylczynski, L\"{u}hmann, and
  Steglich}}]{radu05}
\bibinfo{author}{\bibfnamefont{T.}~\bibnamefont{Radu}},
  \bibinfo{author}{\bibfnamefont{H.}~\bibnamefont{Wilhelm}},
  \bibinfo{author}{\bibnamefont{V.Yushankhai}},
  \bibinfo{author}{\bibfnamefont{D.}~\bibnamefont{Kovrizhin}},
  \bibinfo{author}{\bibfnamefont{R.}~\bibnamefont{Coldea}},
  \bibinfo{author}{\bibfnamefont{Z.}~\bibnamefont{Tylczynski}},
  \bibinfo{author}{\bibfnamefont{T.}~\bibnamefont{L\"{u}hmann}},
  \bibnamefont{and} \bibinfo{author}{\bibfnamefont{F.}~\bibnamefont{Steglich}},
  \bibinfo{journal}{Phys. Rev. Lett.} \textbf{\bibinfo{volume}{95}},
  \bibinfo{pages}{127202} (\bibinfo{year}{2005}).

\bibitem[{\citenamefont{Kosterlitz and Thouless}(1973)}]{kosterlitz73}
\bibinfo{author}{\bibfnamefont{J.~M.} \bibnamefont{Kosterlitz}}
  \bibnamefont{and} \bibinfo{author}{\bibfnamefont{D.~J.}
  \bibnamefont{Thouless}}, \bibinfo{journal}{J. Phys. C: Solid State Phys.}
  \textbf{\bibinfo{volume}{6}}, \bibinfo{pages}{1181} (\bibinfo{year}{1973}).

\bibitem[{\citenamefont{Nelson and Kosterlitz}(1977)}]{nelson77}
\bibinfo{author}{\bibfnamefont{D.~R.} \bibnamefont{Nelson}} \bibnamefont{and}
  \bibinfo{author}{\bibfnamefont{J.~M.} \bibnamefont{Kosterlitz}},
  \bibinfo{journal}{Phys. Rev. Lett.} \textbf{\bibinfo{volume}{39}},
  \bibinfo{pages}{1201} (\bibinfo{year}{1977}).

\bibitem[{\citenamefont{Nelson}(1983)}]{nelson83}
\bibinfo{author}{\bibfnamefont{D.~R.} \bibnamefont{Nelson}}, in
  \emph{\bibinfo{booktitle}{Phase Transitions and Critical Phenomena}}, edited
  by \bibinfo{editor}{\bibfnamefont{C.}~\bibnamefont{Domb}} \bibnamefont{and}
  \bibinfo{editor}{\bibfnamefont{J.}~\bibnamefont{Lebowitz}}
  (\bibinfo{publisher}{Academic Press, London}, \bibinfo{year}{1983}),
  vol.~\bibinfo{volume}{7}, chap.~\bibinfo{chapter}{1}, p.~\bibinfo{pages}{2}.

\bibitem[{\citenamefont{Fisher and Hohenberg}(1988)}]{fisher88}
\bibinfo{author}{\bibfnamefont{D.~S.} \bibnamefont{Fisher}} \bibnamefont{and}
  \bibinfo{author}{\bibfnamefont{P.~C.} \bibnamefont{Hohenberg}},
  \bibinfo{journal}{Phys. Rev. B} \textbf{\bibinfo{volume}{37}},
  \bibinfo{pages}{4936} (\bibinfo{year}{1988}).

\bibitem[{\citenamefont{Stoof and Bijlsma}(1993)}]{stoof93}
\bibinfo{author}{\bibfnamefont{H.~T.~C.} \bibnamefont{Stoof}} \bibnamefont{and}
  \bibinfo{author}{\bibfnamefont{M.}~\bibnamefont{Bijlsma}},
  \bibinfo{journal}{Phys. Rev. E} \textbf{\bibinfo{volume}{47}},
  \bibinfo{pages}{939} (\bibinfo{year}{1993}).

\bibitem[{\citenamefont{Pitaevskii and Stringari}(2003)}]{pitaevskii03}
\bibinfo{author}{\bibfnamefont{L.~P.} \bibnamefont{Pitaevskii}}
  \bibnamefont{and}
  \bibinfo{author}{\bibfnamefont{S.}~\bibnamefont{Stringari}},
  \emph{\bibinfo{title}{{B}ose-{E}instein Condensation}}
  (\bibinfo{publisher}{Clarendon Press, Oxford}, \bibinfo{year}{2003}).

\bibitem[{\citenamefont{Takahashi}(1957)}]{takahashi57}
\bibinfo{author}{\bibfnamefont{Y.}~\bibnamefont{Takahashi}},
  \bibinfo{journal}{Nuovo Cimento} \textbf{\bibinfo{volume}{6}},
  \bibinfo{pages}{371} (\bibinfo{year}{1957}).

\bibitem[{\citenamefont{Ward}(1950)}]{ward50}
\bibinfo{author}{\bibfnamefont{J.~C.} \bibnamefont{Ward}},
  \bibinfo{journal}{Phys. Rev.} \textbf{\bibinfo{volume}{78}},
  \bibinfo{pages}{182} (\bibinfo{year}{1950}).

\bibitem[{\citenamefont{Griffin}(1994)}]{griffin94}
\bibinfo{author}{\bibfnamefont{A.}~\bibnamefont{Griffin}},
  \emph{\bibinfo{title}{Excitations in a {B}ose-condensed Liquid}}
  (\bibinfo{publisher}{Cambridge University Press, Cambridge},
  \bibinfo{year}{1994}).

\bibitem[{\citenamefont{Kavokin et~al.}(2003)\citenamefont{Kavokin, Malpuech,
  and Laussy}}]{kavokin03}
\bibinfo{author}{\bibfnamefont{A.}~\bibnamefont{Kavokin}},
  \bibinfo{author}{\bibfnamefont{G.}~\bibnamefont{Malpuech}}, \bibnamefont{and}
  \bibinfo{author}{\bibfnamefont{F.~P.} \bibnamefont{Laussy}},
  \bibinfo{journal}{Phys. Lett. A} \textbf{\bibinfo{volume}{306}},
  \bibinfo{pages}{187} (\bibinfo{year}{2003}).

\bibitem[{\citenamefont{Malpuech et~al.}(2003)\citenamefont{Malpuech, Rubo,
  Laussy, Bigenwald, and Kavokin}}]{malpuech03}
\bibinfo{author}{\bibfnamefont{G.}~\bibnamefont{Malpuech}},
  \bibinfo{author}{\bibfnamefont{Y.~G.} \bibnamefont{Rubo}},
  \bibinfo{author}{\bibfnamefont{F.~P.} \bibnamefont{Laussy}},
  \bibinfo{author}{\bibfnamefont{P.}~\bibnamefont{Bigenwald}},
  \bibnamefont{and} \bibinfo{author}{\bibfnamefont{A.~V.}
  \bibnamefont{Kavokin}}, \bibinfo{journal}{Semicond. Sci. Technol.}
  \textbf{\bibinfo{volume}{18}}, \bibinfo{pages}{S395} (\bibinfo{year}{2003}).

\bibitem[{\citenamefont{Nagaosa}(1999)}]{nagaosa}
\bibinfo{author}{\bibfnamefont{N.}~\bibnamefont{Nagaosa}},
  \emph{\bibinfo{title}{Qauntum Field Theory in Condensed Matter Physics}}
  (\bibinfo{publisher}{Springer-Verlag, Berlin}, \bibinfo{year}{1999}).

\bibitem[{\citenamefont{Keeling et~al.}(2005)\citenamefont{Keeling, Eastham,
  Szymanska, and Littlewood}}]{keeling05}
\bibinfo{author}{\bibfnamefont{J.}~\bibnamefont{Keeling}},
  \bibinfo{author}{\bibfnamefont{P.~R.} \bibnamefont{Eastham}},
  \bibinfo{author}{\bibfnamefont{M.~H.} \bibnamefont{Szymanska}},
  \bibnamefont{and} \bibinfo{author}{\bibfnamefont{P.~B.}
  \bibnamefont{Littlewood}}, \bibinfo{journal}{Phys. Rev. B}
  \textbf{\bibinfo{volume}{72}}, \bibinfo{pages}{115320}
  (\bibinfo{year}{2005}).

\bibitem[{\citenamefont{Khawaja et~al.}(2002)\citenamefont{Khawaja, Andersen,
  Proukakis, and Stoof}}]{khawaja02}
\bibinfo{author}{\bibfnamefont{U.~A.} \bibnamefont{Khawaja}},
  \bibinfo{author}{\bibfnamefont{J.~O.} \bibnamefont{Andersen}},
  \bibinfo{author}{\bibfnamefont{N.~P.} \bibnamefont{Proukakis}},
  \bibnamefont{and} \bibinfo{author}{\bibfnamefont{H.~T.~C.}
  \bibnamefont{Stoof}}, \bibinfo{journal}{Phys. Rev. A}
  \textbf{\bibinfo{volume}{66}}, \bibinfo{pages}{013615}
  (\bibinfo{year}{2002}).

\end{thebibliography}
\end{document}